\documentclass{jpp}
\usepackage{graphicx}
\usepackage{amsmath,amssymb}
\usepackage{natbib}
\usepackage[hidelinks]{hyperref}
\usepackage[retain-unity-mantissa=false]{siunitx}
\usepackage{etextools}
\usepackage{color}
\newcommand{\diff}{\mathrm d}
\renewcommand{\vec}[1]{\mathbf{#1}}

\title[Pair production in bi-frequent fields]{Pair production by Schwinger and Breit-Wheeler processes
in bi-frequent fields}

\author[A.~Otto, T.~Nousch et al.]{A.~Otto$^1$\thanks{Email address for
correspondence: a.otto@hzdr.de}, T.~Nousch$^1$, D.~Seipt$^2$, B.~K\"ampfer$^1$,\\
D.~Blaschke$^{3,4,5}$, A.~D.~Panferov$^6$, S.~A.~Smolyansky$^6$, A.~I.~Titov$^{4}$}

\affiliation{1 Institute of Radiation Physics, Helmholtz-Zentrum
Dresden-Rossendorf,\\ Bautzner Landstra\ss e 400, 01328 Dresden, Germany\\
Institut f\"ur Theoretische Physik, Technische Universit\"at Dresden,\\
Zellescher Weg 17, 01062 Dresden, Germany\\[\affilskip]
2 Helmholtz-Institut Jena, Fr\"obelstieg 3, 07743 Jena, Germany\\
Theoretisch-Physikalisches Institut, Friedrich-Schiller-Universit\"at Jena,\\
Max-Wien-Platz 1, 07743 Jena, Germany\\[\affilskip]
3 Institute for Theoretical Physics, University of Wroclaw,\\ pl.\ M.\ Borna 9,
50-204 Wroclaw, Poland\\[\affilskip]
4 Bogoliubov Laboratory for Theoretical Physics, JINR Dubna,\\ Joliot-Curie
str.\ 6, 141980 Dubna, Russia\\[\affilskip]
5 National Research Nuclear University (MEPhI), Kashirskoe Shosse 31,\\
115409 Moscow, Russia\\[\affilskip]
6 Department of Physics, Saratov State University, 410071 Saratov,
Russia}

\begin{document}

\maketitle

\begin{abstract}
Counter-propagating and suitably polarized light (laser) beams can provide conditions for pair production.
Here, we consider in more detail the following two situations:
(i) In the homogeneity regions of anti-nodes of linearly polarized ultra-high
intensity laser beams, the Schwinger
process is dynamically assisted by a second high-frequency field, e.g. by a XFEL beam.
(ii) A high-energy probe photon beam colliding with a superposition of co-propagating  intense laser and
XFEL beams gives rise to the laser assisted Breit-Wheeler process. Prospects of such bi-frequent field constellations with
respect to the feasibility of conversion of light into matter are discussed.
\end{abstract}

\section{Introduction}

The Schwinger effect~\citep{sauter,schwinger} means the instability of a spatially homogeneous,
purely electric field with respect to the decay into a state with pairs, e.g.
electrons ($e^-$) and positrons ($e^+$), and a screened electric field,
symbolically $\vert \vec E \rangle \to \vert \vec E^\prime e^+ e^- \rangle$
(cf.~\citep{gelis_schwinger_2015} for a recent review).
The pair creation rate $w \propto \exp\{ - \pi E_c / \vert \vec E \vert \}$ for fields
attainable presently in mesoscopic laboratory installations is exceedingly small since the
Sauter-Schwinger (critical) field strength
$E_c=m^2/\vert e\vert=\SI{1.3e18}{V\per m}$ is for electrons/positrons with
masses $m$ and charges $\pm e$ so large (we employ here natural units with $c = \hbar = 1$).
The notion of dynamical Schwinger process refers to a situation where the spatially homogeneous
electric field has a time dependence, $\vec E(t)$. The particular case of a periodic field
is dealt with in~\citep{brezin_pair_1970} with the motivation that tightly focused laser beams
can provide high field strengths, e.g.\ in the anti-nodes of pair-wise counter
propagating, linearly polarized beams.
The superposition of many laser beams, as considered, e.g.\
in~\citep{narozhny_pair_2004}, can enlarge the pair yield noticeably. A particular variant is the
superposition of strong laser beams and weaker but high-frequency beams which
may be idealized as a common classical background field
$\vec E(t) = \vec E_1(\omega t) + \vec E_2 (N \omega t)$.
If the frequency of the second field, $N \omega$ is sufficiently large, the
tunneling path through the
positron-electron gap is shortened by the assistance of the multi-photon effect~\citep{schutzhold_dynamically_2008,dunne_catalysis_2009} and, as a consequence,
the pair production is enhanced.
This dynamically assisted Schwinger process supposes a Keldysh parameter
$\gamma_1 = (E_c / E_1) (\omega/m) \ll 1$ to stay in the tunneling
regime\footnote[2]{Similar to ionization in atomic physics,
one can also for pair production distinguish between a tunneling ($\gamma\ll1$)
and a multi-photon regime ($\gamma\gg1$), depending on the value of the Keldysh
parameter $\gamma$.}.
The combination $\gamma_1 < 1$ and $\gamma_2 = (E_c / E_2 ) (N\omega/m) > 1$
is dubbed assisted dynamical Schwinger effect since the field ``1''
with parameters $E_1$, $\omega$ refers to the dynamical Schwinger effect in the
nomenclature of~\citep{brezin_pair_1970}, and the field ``2'' with parameters
$E_2$, $N\omega$ is assisting.
Various pulse shapes for $E_{1,2}$ have been
studied with the goal to seek for optimal combinations~\citep{hebenstreit_optimization_2014,kohlfurst_optimizing_2013,akal_electron-positron_2014}.
Current lasers reach intensities of $\SI{2e22}{W\per cm^2}$
(cf.~\citep{di_piazza_extremely_2012} for an overview) corresponding to an
inverse Keldysh parameter of $\gamma^{-1}=10$.
Planned facilities are, for example, ELI-NP~\citep{eli} and Apollon~\citep{zou_design_2015} ($\SI{10}{PW}$,
$\SI{1e22}{W\per cm^2}$) or HiPER~\citep{hiper} ($\SI{100}{PW}$, $\SI{1e26}{W\per cm^2}$).
(The Sauter-Schwinger field strength requires an intensity of $\SI{4e29}{W\per cm^2}$.)

All these investigations aim at verifying the decay of the vacuum. Besides the mentioned strong (but
presently not strong enough) fields also the Coulomb fields accompanying heavy and
super-heavy atomic nuclei
have been considered as an option to study the vacuum break down~\citep{greiner_3,rafelski_superheavy_1971,muller_solution_1972,muller_solution_1973,bialynicki-birula_phase-space_1991}.
Previous experiments, however, have not been
conclusive~\citep{heinz_positron_2000}.

Another avenue for pair creation is the conversion of light into matter in the collision of
photon beams. The Breit-Wheeler process~\citep{breit_wheeler} refers
to the reaction
$\gamma^\prime + \gamma \to e^+ + e^-$
which is a crossing channel of the Compton process or the time-reversed annihilation.
The famous experiment E-144 at SLAC~\citep{burke_positron_1997} can be interpreted as a two-step
process with Compton backscattering of a laser beam and subsequent reaction
of the Compton backscattered photons with
the laser beam in non-linear Breit-Wheeler pair production~\citep{burke_positron_1997,bamber_studies_1999}. The notion non-linear
Breit-Wheeler process means the instantaneous reaction with a multiple of laser beam photons,
i.e.\ $\gamma^\prime + n \omega_L \to e^+ + e^-$.
Also here one
can ask whether the laser assisted non-linear Breit-Wheeler process
$\gamma^\prime + \omega_{XFEL} + n \omega_L \to e^+ + e^-$ shows peculiarities
due to the superposition of the co-propagating XFEL and laser beams.

Other field combinations, such as the nuclear Coulomb field and XFEL/laser beams,
are also conceivable~\citep{augustin_nonlinear_2014,di_piazza_effect_2010} (cf.~\citep
{di_piazza_extremely_2012} for a recent review and further references),
but will not be addressed here.

Our paper is organized as follows. In section 2 we consider the reasoning for forming
resonance type structures in the phase space distribution of pairs created in the
assisted dynamical Schwinger process. The considered classical
background field configuration has been characterized above:
the superposition of two spatially homogeneous fields of different strengths and
frequencies with a common envelope, as investigated in~\citep{otto_lifting_2015,
otto_dynamical_2015,panferov_assisted_2015}.
Examples are given for the mutual amplification, and some glimpses on the time
evolution in simple pulses are provided too.
Section~3 deals with the laser assisted Breit-Wheeler process, where spectral
caustics have identified already in~\citep{nousch_spectral_2016}.
Specifically, we show here the sensitivity of the spectral caustics on the laser
beam intensity which is important for multi-shot experiments with not perfectly
tuneable intensity parameter.
Our approach here utilizes the common XFEL + laser field again as a
classical background field to be dealt with in the Furry picture, while the
probe photon $\gamma'$ refers to a quantized radiation field.
We briefly summarize in Section~4.

\section{Assisted dynamical Schwinger process}
In this section we consider pair production in the spirit of the Schwinger
process, i.e.\ creation of $e^\pm$ pairs by a purely electric background field
which is assumed to be spatially homogeneous.
Int the following, we use the notation and formalism as introduced in~\citep
{otto_lifting_2015}.
The quantum kinetic equation~\citep{schmidt_quantum_1998}
\begin{equation}\label{QKE}
\dot f(\vec p ,t) = \frac{\lambda(\vec p, t)}{2}
\int\limits^t_{- \infty} \diff t^{\prime}
\lambda(\vec p, t^{\prime}) (1 - 2 f(\vec p, t^{\prime}))
\cos\theta(\vec p, t, t^{\prime})
\end{equation}
determines the time $(t)$ evolution of the dimensionless phase space
distribution function per spin projection degree of freedom\footnote[5]{
In~\citep{otto_lifting_2015,otto_dynamical_2015} we employ a different
convention with a sum over spin degrees of freedom, i.e.\ $f\to\sum_sf$ which
removes factors $2$ in front of $f$.}
$f(\vec p, t) = \diff N(\vec p, t) / \diff^3p \, \diff^3x$, where
$N$ refers to the particle number and $\diff^3p$ and $\diff^3x$ are the three
dimensional volume elements in momentum ($p$) and configuration ($x$) spaces.
We emphasize that only $f(\vec p, t \to +\infty)$ can be considered as single
particle distribution which may represent the source term of a subsequent time
evolution of the emerging $e^+e^-$ plasma.
The initial condition for solving (\ref{QKE}) is $f(\vec p, t \to -\infty) = 0$.
Screening and backreaction are not included with virtue of the small values of $f$ in subcritical fields (cf.~\citep{gelis_formulation_2013} for recent work on
that issue).
Above the quantities $\lambda(\vec p, t) = \frac{eE(t)\,
\varepsilon_\perp(p_\perp)}{\varepsilon^2(\vec p, t)}$ stand for the amplitude
of the vacuum transition, and $\theta(\vec p, t, t') = 2\int^t_{t'}\diff\tau\,
\varepsilon(\vec p, \tau)$ for the dynamical phase, describing the vacuum
oscillations modulated by the external field;
the quasi-energy $\varepsilon$, the transverse energy $\varepsilon_\perp$
and the longitudinal quasi-momentum $P$ are defined as
$\varepsilon(\vec p, t) = \sqrt{\varepsilon_\perp^2 (p_\perp) + P^2(p_\parallel, t)}$
and
$\varepsilon_\perp (p_\perp)= \sqrt{m^2 + p^2_\perp}, $
$ P(p_\parallel, t) = p_\parallel -eA(t), $
where $p_\perp=|\vec p_\perp|$ is the modulus of the kinetic
momentum ($\vec p$) component of positrons (electrons)
perpendicular to the electric field, and $p_\parallel$ denotes the $E$-parallel
kinetic momentum component.
The electric field follows from the potential
\begin{equation}
A = K(\omega t) \left( \frac{E_1}{\omega} \cos (\omega t)
+ \frac{E_2}{N \omega} \cos (N \omega t) \right)
\label{A_AO}
\end{equation}
by $E = - \dot A$ in Coulomb gauge. Equation (\ref{A_AO}) describes a bi-frequent
field with frequency ratio $N$ (integer) and field strengths $E_1$ -- the strong field ``1'' --
and $E_2$ -- the weak field ``2''. The quantity $K$ is the common envelope
function with the properties (i) absolutely flat in the flat-top time interval
$-t_\text{f.t.}/2 < t < + t_\text{f.t.}/2$ and (ii) absolutely zero for $t < - t_\text{f.t.}/2 - t_\text{ramp}$
and $t > t_\text{f.t.}/2 + t_\text{ramp}$ and (iii) absolutely smooth everywhere, i.e.\
$K$ belongs to the $C^\infty$ class; $t_\text{ramp}$ is the ramping duration
characterizing the switching on/off time intervals.

\begin{figure}
\centering
\includegraphics[width=\textwidth]{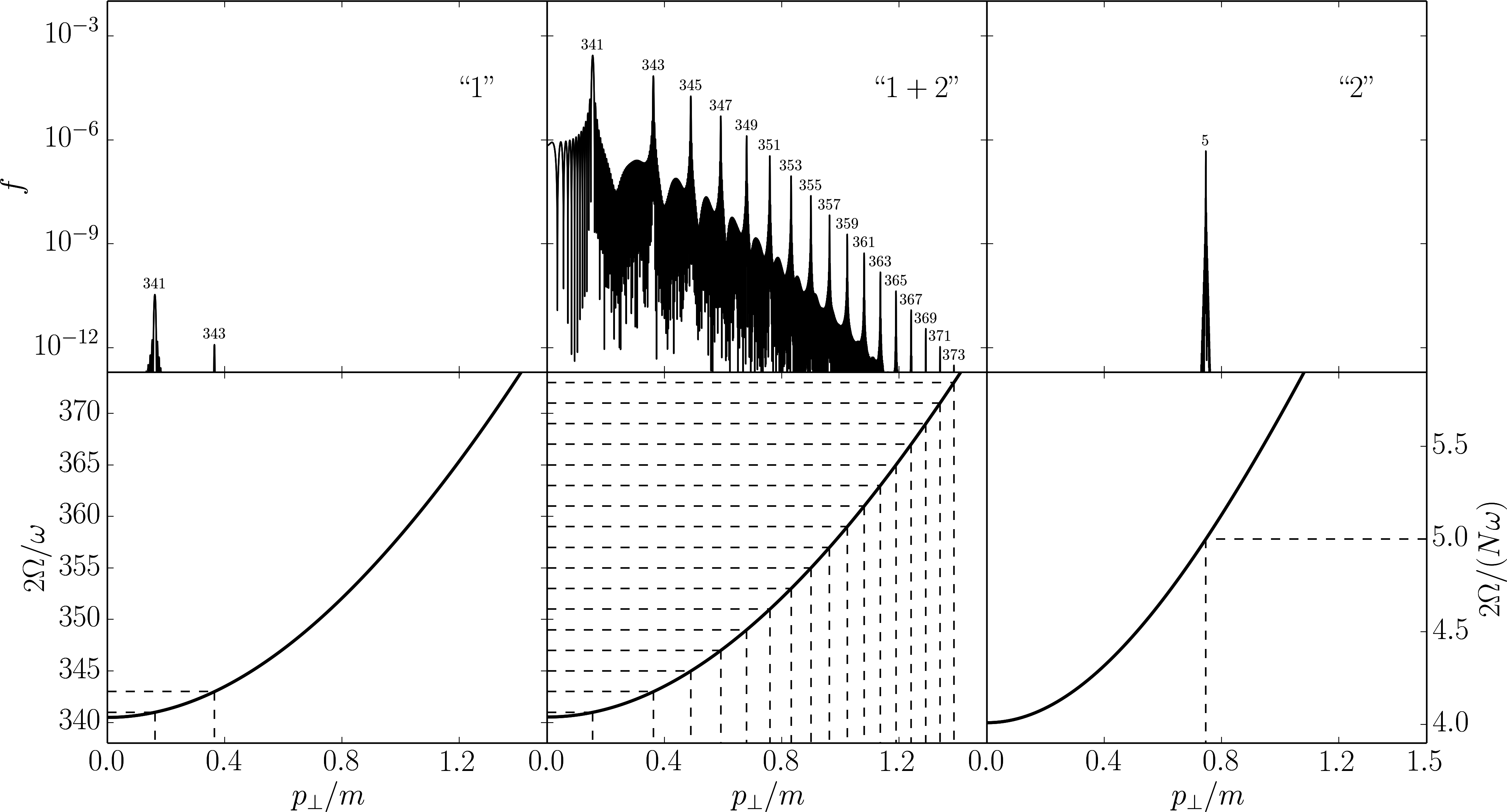}
\caption{Top row: Asymptotic transverse momentum ($p_\perp$) spectrum
at $p_\parallel=0$ for the bi-frequent field (\ref{A_AO}) (middle panel) and the
field components ``1'' (left panel, $E_1=0.1\,E_c$, $\omega=0.02\,m$) and
``2'' (right panel, $E_2=0.05\,E_c$, $N=25$) alone.
Bottom row: Fourier zero-modes $2\Omega(p_\perp, p_\parallel=0)$ scaled by
$\omega$ (left and middle panels) and $N\omega$ (right panel) for the fields
in the top row with resonance conditions (horizontal dashed
lines for $\ell=341$ and $343$ (left; higher-$\ell$ resonances are not depicted
since the peaks are underneath the scale displayed in the top panel), $\ell=341,
\dots,373$ (middle) and $\ell=5$ (right);
vertical dashed lines are for the resonance positions;
peaks for even $\ell$ appear only for $p_\parallel\ne0$ but get a zero amplitude
at $p_\parallel=0$, and thus their positions are not depicted).}
\label{fig:1AO}
\end{figure}

Figure \ref{fig:1AO} (top row) exhibits three examples for the transverse phase
space distribution $f(p_\perp,p_\parallel=0,t\to\infty)$ for
$E_1=0.1\,E_c$, $E_2=0.05\,E_c$, $\omega=0.02\,m$, $N=25$,
$t_\text{ramp}=5\,\omega^{-1}$ and $t_\text{f.t.}=25\,\omega^{-1}$
obtained by numerically solving Eq.~\eqref{QKE}.
The chosen parameters are by far not yet in reach at present and near-future
facilities.
Due to the periodicity of the involved fields
and their finite duration a pronounced peak structure emerges (the peaks become
sharp, elliptically bend ridges with deep notches when continuing the spectrum
to finite values of  $p_\parallel$).
The peak heights scale with $t_\text{f.t.}^2$ for not too long pulse
duration.
The peak positions are determined by the resonance condition~\citep
{otto_lifting_2015}
\begin{equation}
2 \Omega (p_\perp, p_\parallel) - \ell \omega = 0,
\label{resonance}
\end{equation}
where
$\Omega = \frac{m}{2\pi} \int_0^{2 \pi} \diff x
\sqrt{1 + (p_\perp / m)^2 +
[(p_\parallel / m) - \gamma_1^{-1} \cos x - \gamma_2^{-1} \cos N x ]^2}$
is the Fourier zero-mode of $\varepsilon$.
The values of $\ell$ (integer) where the resonance condition (\ref{resonance})
is fulfilled can be used to label the peaks. $\Omega (p_\perp = p_\parallel = 0)$
may be interpreted as effective mass $m^*$~\citep{kohlfurst_effective_2014} which determines
$\ell_\text{min} = int(1+2m^*/\omega)$.
The Fourier zero-modes as functions of $p_\perp$ at $p_\parallel=0$ are
displayed in the bottom row in Fig.~\ref{fig:1AO} together with the
resonance positions.
For the field ``1'' alone (left bottom panel) one has to take the limit
$\gamma_2\to\infty$ in the Fourier zero-mode, while field ``2'' alone (right
bottom panel) corresponds to $\gamma_1\to\infty$ and the replacement
$\omega\to N\omega$ in~\eqref{resonance}.

The striking feature in Fig.~\ref{fig:1AO} (cf.~\citep{otto_lifting_2015,
otto_dynamical_2015} for other examples with different parameters, in particular
$t_\text{f.t.}$, and~\citep{haehnel_bachelor_2015} for a wider range of
field strengths) is the lifting of the spectrum related
to field ``1'' by the assistance of field ``2''. While the amplification of the created
pair distribution by the assistance field can be huge, for sub-critical fields
the frequency $N \omega$ must be ${\cal O} (m)$ to overcome the exponential
suppression. This implies that intensities envisaged in ELI pillar
IV~\citep{eli} must be at our disposal in conjunction with much higher
frequencies to arrive at measurable pair numbers enhanced further by an
assistant field~\citep{otto_dynamical_2015}.

\begin{figure}
\centering
\includegraphics[width=\textwidth]{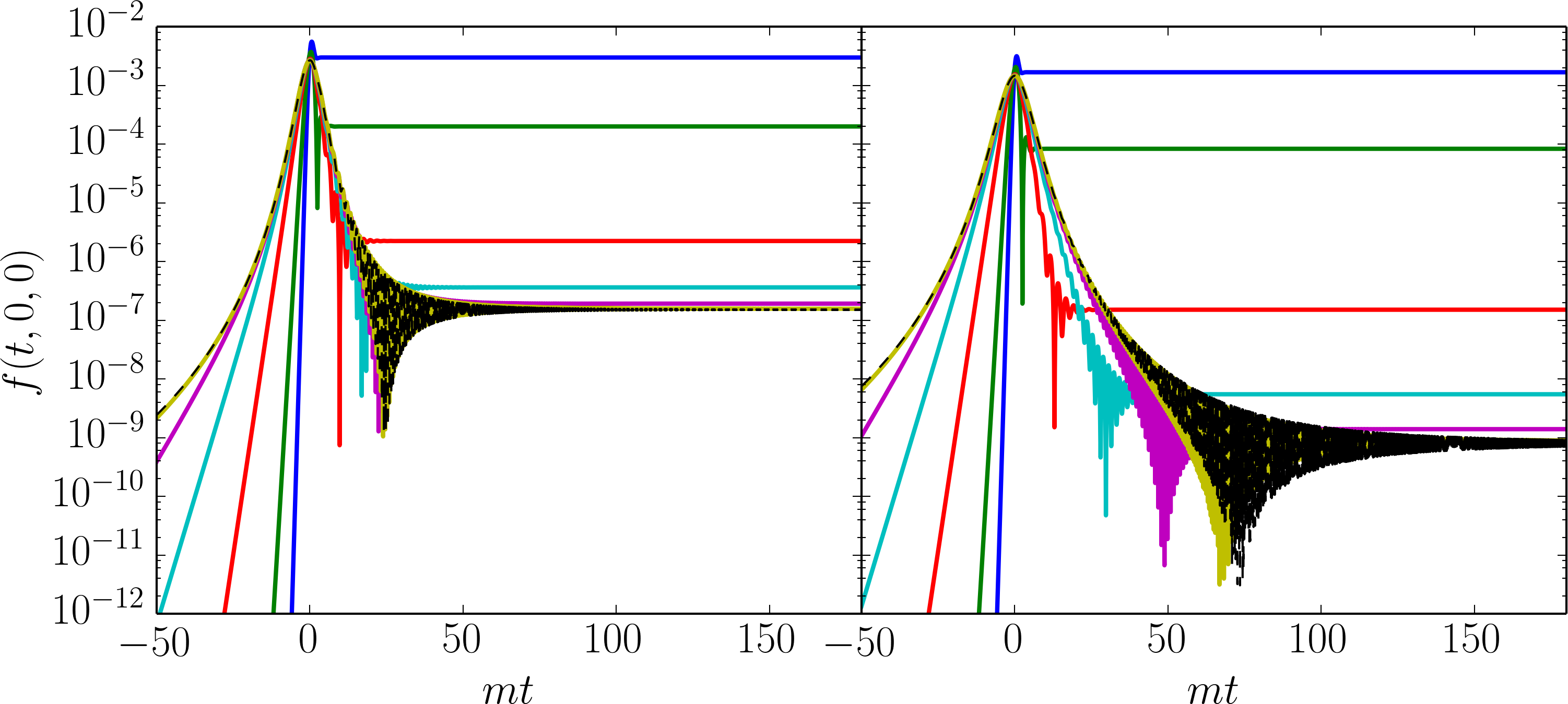}
\caption{Time evolution of $f(p_\perp = p_\parallel = 0, t)$ in the adiabatic
basis for the Sauter pulse~\eqref{sauter} for $\tau=1\,m^{-1}$ (blue),
$\tau=2\,m^{-1}$ (green), $\tau=5\,m^{-1}$ (red), $\tau=10\,m^{-1}$ (cyan),
$\tau=20\,m^{-1}$ (purple), $\tau=50\,m^{-1}$ (yellow) and $E_0=0.2\,E_c$
(left panel), $E_0=0.15\,E_c$ (right panel).
The dashed black curves depict the Schwinger case as the limit of large values
of $\tau$.
Note the vast drop of the residual phase space occupancy for larger values of
$\tau$ when changing $E_0$ from $0.2\,E_c$ to $0.15\,E_c$.}
\label{fig:2AO}
\end{figure}

\begin{figure}
\centering
\includegraphics[width=0.8\textwidth]{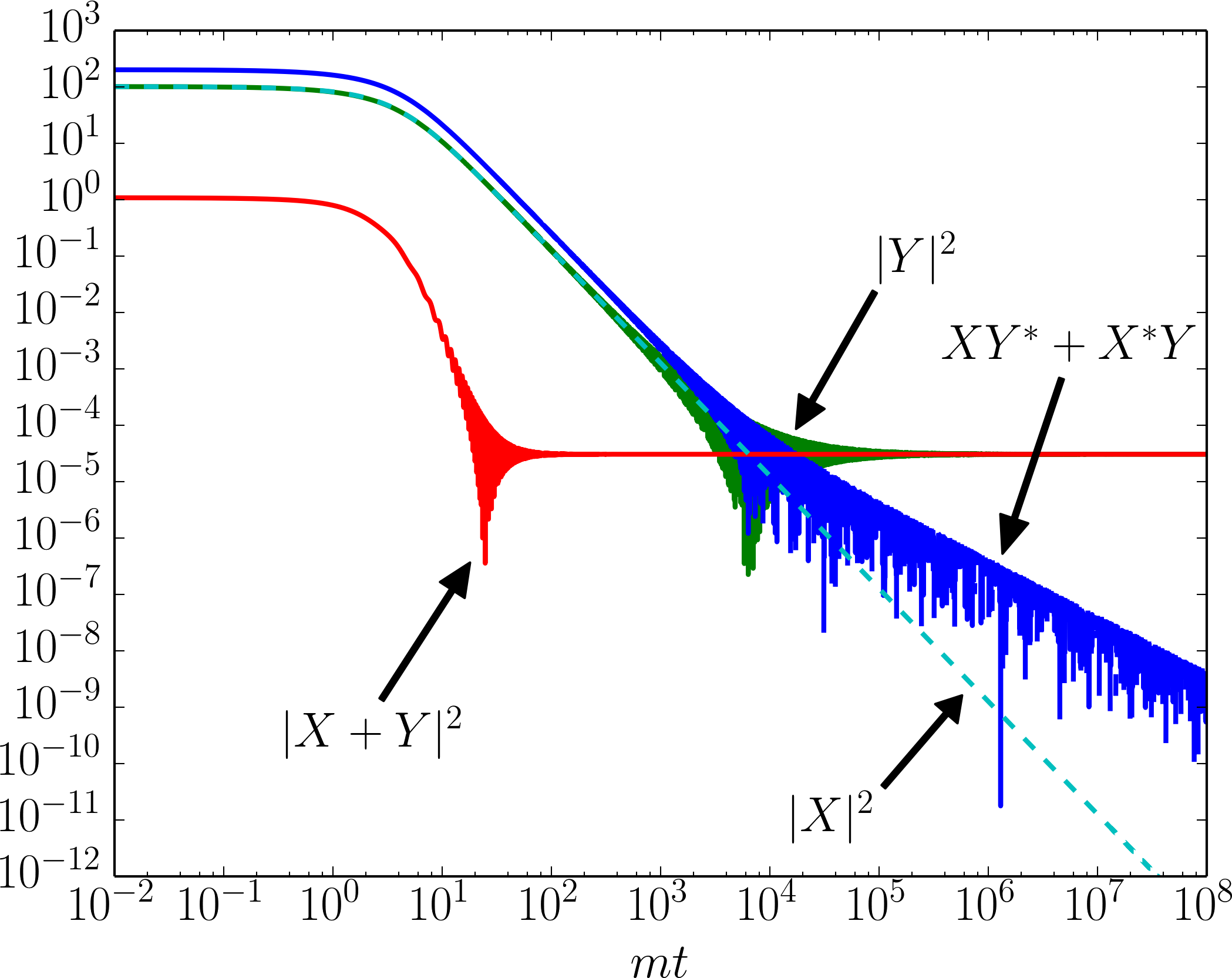}
\caption{Time evolution of the components defined in~\eqref{components} of the analytical solution~\eqref{f_components} of the Schwinger case depicted for
$E_0=0.2\,E_c$.
Cyan dashed curve: $\vert X \vert^2$,
green curve: $\vert Y \vert^2$,
blue curve: interference term $X Y^* + X^* Y$,
red curve:  $\vert X + Y \vert^2$. }
\label{fig:3AO}
\end{figure}

Even with low pair creation probability a once produced pair may seed a further
avalanche evolution~\citep{bell_possibility_2008,king_photon_2013,
elkina_qed_2011} toward an electron-positron plasma.
In this respect one may ask for the time scales to approach
the asymptotic out-state. A unique answer seems not to be achievable within
the present framework due to the
unavoidable ambiguity of the particle definition (see, e.g.~\citep{dabrowski_super-adiabatic_2014} for
examples of changing the time evolution of $f$ at intermediate times
when changing the basis). Having this disclaimer in mind one can inspect
nevertheless graphs of $f(t)$. Figure \ref{fig:2AO} exhibits the time evolution
in the adiabatic basis for the Sauter pulse
\begin{align}
E(t) = \frac{E_0}{\cosh^2 (t/\tau)}\:.
\label{sauter}
\end{align}
which is fairly different from~\eqref{A_AO}.
The analytical solution~\citep{narozhny_the_1970,hebenstreit_diss_2011} of
equation (\ref{QKE}) is useful for checking numerical codes which are challenged
by dealing with rapidly changing functions over many orders of magnitude.
For large values of the pulse duration parameter $\tau$ the Schwinger case is
recovered, see~\citep{hebenstreit_diss_2011}:
\begin{equation}
f = \frac{1}{8}\left(1+\frac{u}{\sqrt{2\hat\eta+u^2}}\right)
\mathrm e^{-\frac{\pi\hat\eta}{4}} \vert X + Y \vert^2
\label{f_components}
\end{equation}
with
\begin{align}
X = \left(\sqrt{2\hat\eta+u^2}-u\right)D_{-1+\frac{i\hat\eta}{2}}
\left(-u\mathrm e^{-\frac{i\pi}{4}}\right)\:,\quad
Y = -2\mathrm e^{\frac{i\pi}{4}}D_{\frac{i\hat\eta}{2}}
\left(-u\mathrm e^{-\frac{i\pi}{4}}\right)\:,
\label{components}
\end{align}
where $D$ is the parabolic cylinder function, $u=\sqrt{\frac{2}{|e|E_0}}
(p_\parallel+eE_0t)$ and $\hat\eta=\frac{m^2+p_\perp^2}{|e|E_0}$.
While for $E = 0.2 E_c$ the net function $\propto \vert X + Y \vert^2$ reaches its
asymptotic value already at $t m \approx 20$ (see Fig.~\ref{fig:3AO}),
the individual components
$\vert X \vert^2$, $\vert Y \vert^2$ and $X Y^* + X^* Y$ display a violent
time dependence on much longer times. Note also the subtle cancellations.

In the case of the Sauter pulse, see Fig.~\ref{fig:2AO}, the asymptotic
values of $f$ are reached at shorter times with decreasing values of $\tau$.
The relatively large values of $f(t\approx 0)$ have tempted sometimes researchers
to relate them to particular effects caused by the transient state. Clearly, only
observables, e.g. provided by probe beams, at asymptotic times are reliable.
It is questionable, however, whether such probes can disentangle transient state
contributions and asymptotic state contributions in a unique manner.

\section{Laser assisted Breit-Wheeler process}
\begin{figure}
\centering
\includegraphics[width=\textwidth]{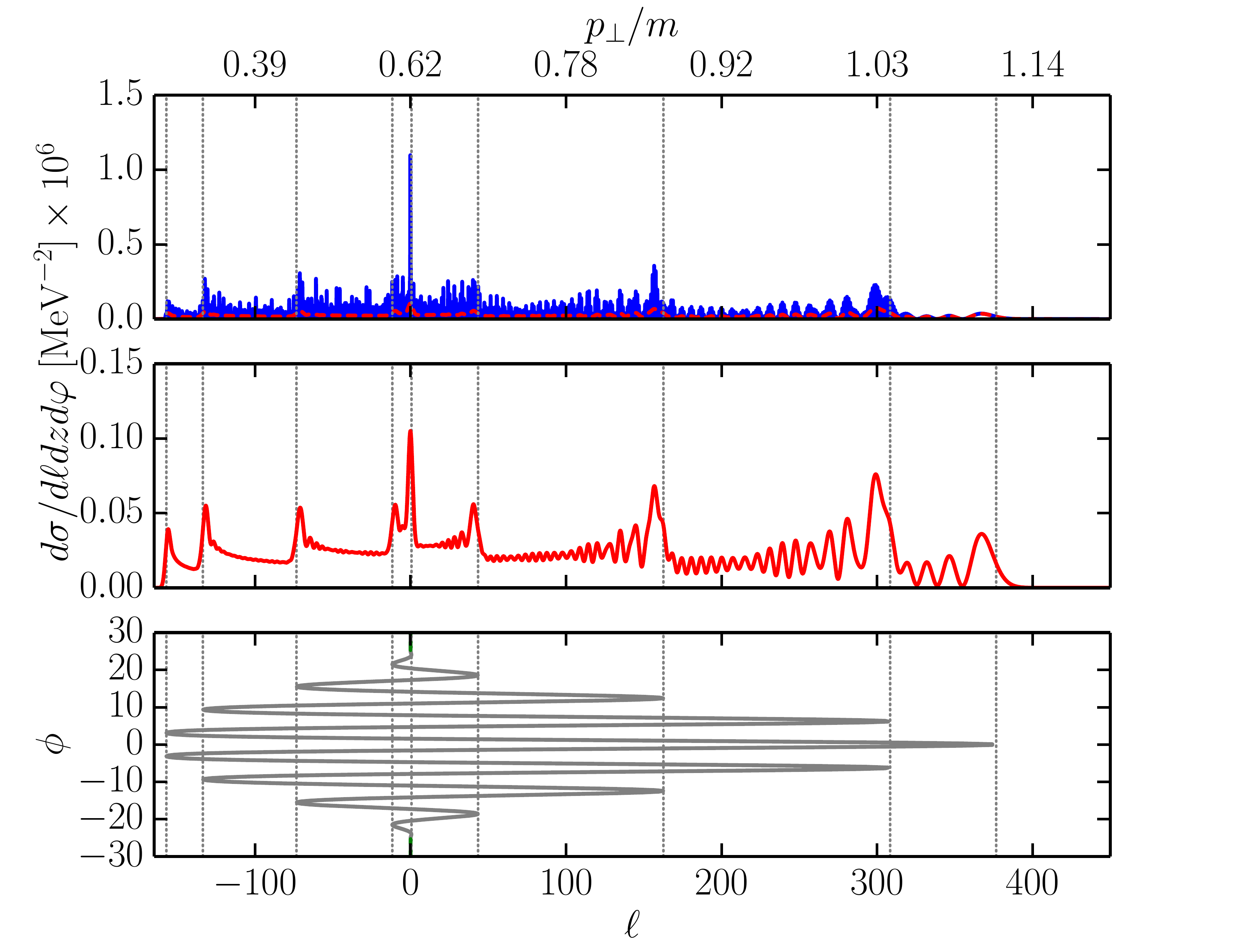}
\caption{Spectra for the laser assisted Breit-Wheeler process for a probe photon
of energy $\SI{60}{MeV}$ colliding head-on with an XFEL photon (energy
$\SI{6}{keV}$) and a co-propagating laser beam (frequency $\SI{10}{eV}$).
Further parameters are $\eta=1/600$, $\gamma_X=10^5$,
$\tau_X=7\tau/(4\pi\eta)$, $\gamma_L=2$ and $\tau_L=8\pi$ in the
field~\eqref{A_Tobias}.
These parameters translate into intensities of
$\SI{6.2e15}{W\per cm^2}$ and $\SI{4.3e19}{W\per cm^2}$ for XFEL and laser,
respectively.
Upper panel: $\diff\sigma/\diff\ell\diff z\diff\varphi$ at $z=0$ and
$\varphi=\pi$ as a function of $\ell$ (lower axis; the corresponding values of
$p_\perp$ are given at the upper axis).
The calculated spectrum is smoothed by a Gaussian window function with width
$\delta=1.3$ to get the red curve.
Middle panel: smoothed spectrum separately.
Lower panel: phase $\phi$ as a function of $\ell$ (see~\citep
{nousch_spectral_2016} for details).
The vertical dotted lines depict the positions of diverging
$\diff\phi/\diff\ell$, where two branches of $\phi(\ell)$ merge.}
\label{fig:1TN}
\end{figure}
\begin{figure}
\centering
\includegraphics[width=\textwidth]{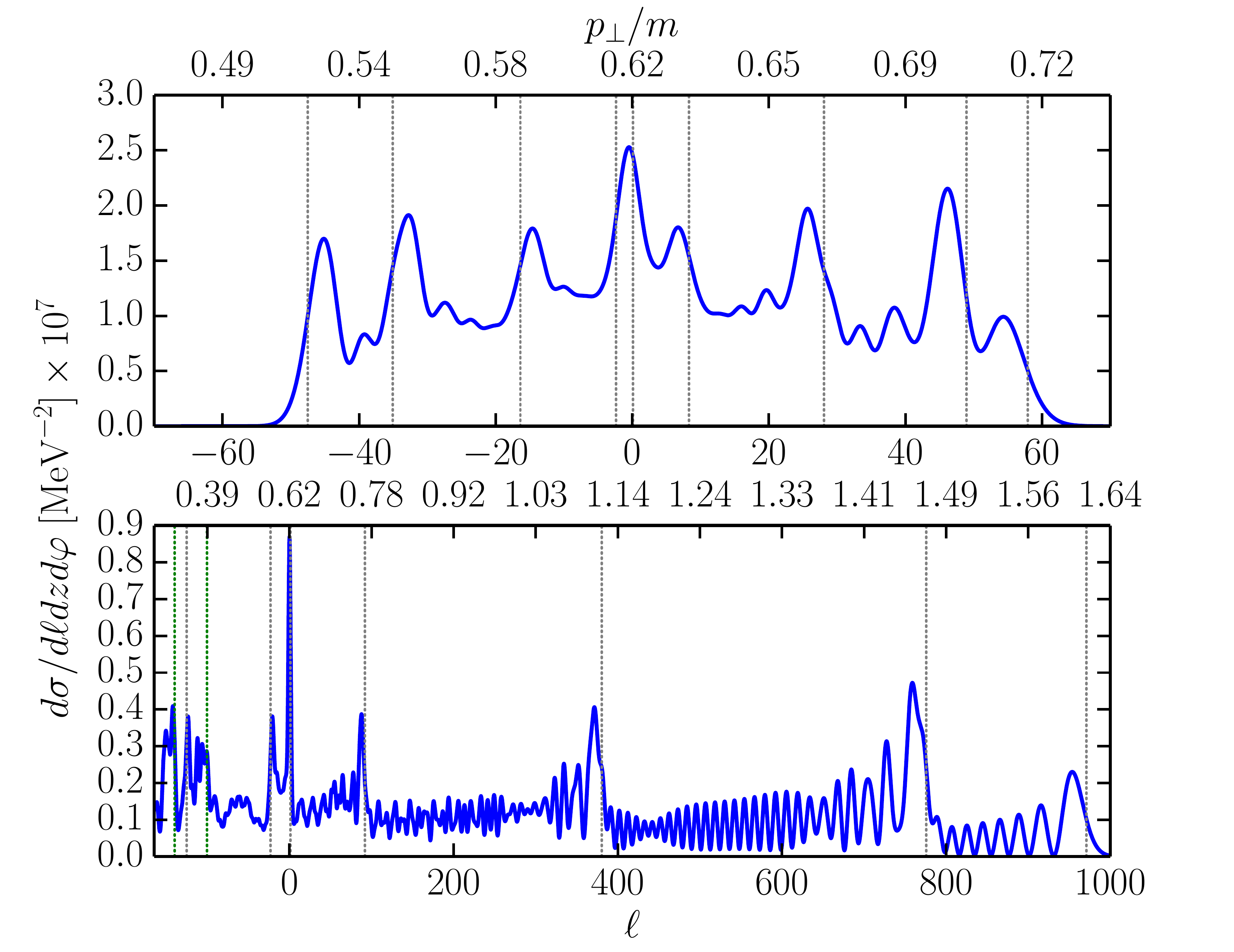}
\caption{As middle panel in Fig.~\ref{fig:1TN} but for $\gamma_L=10$,
laser intensity $\SI{1.7e18}{W\per cm^2}$ (top panel) and
$\gamma_L=1$, laser intensity $\SI{1.7e20}{W\per cm^2}$ (bottom
panel).}
\label{fig:3TN}
\end{figure}
\begin{figure}
\centering
\includegraphics[width=\textwidth]{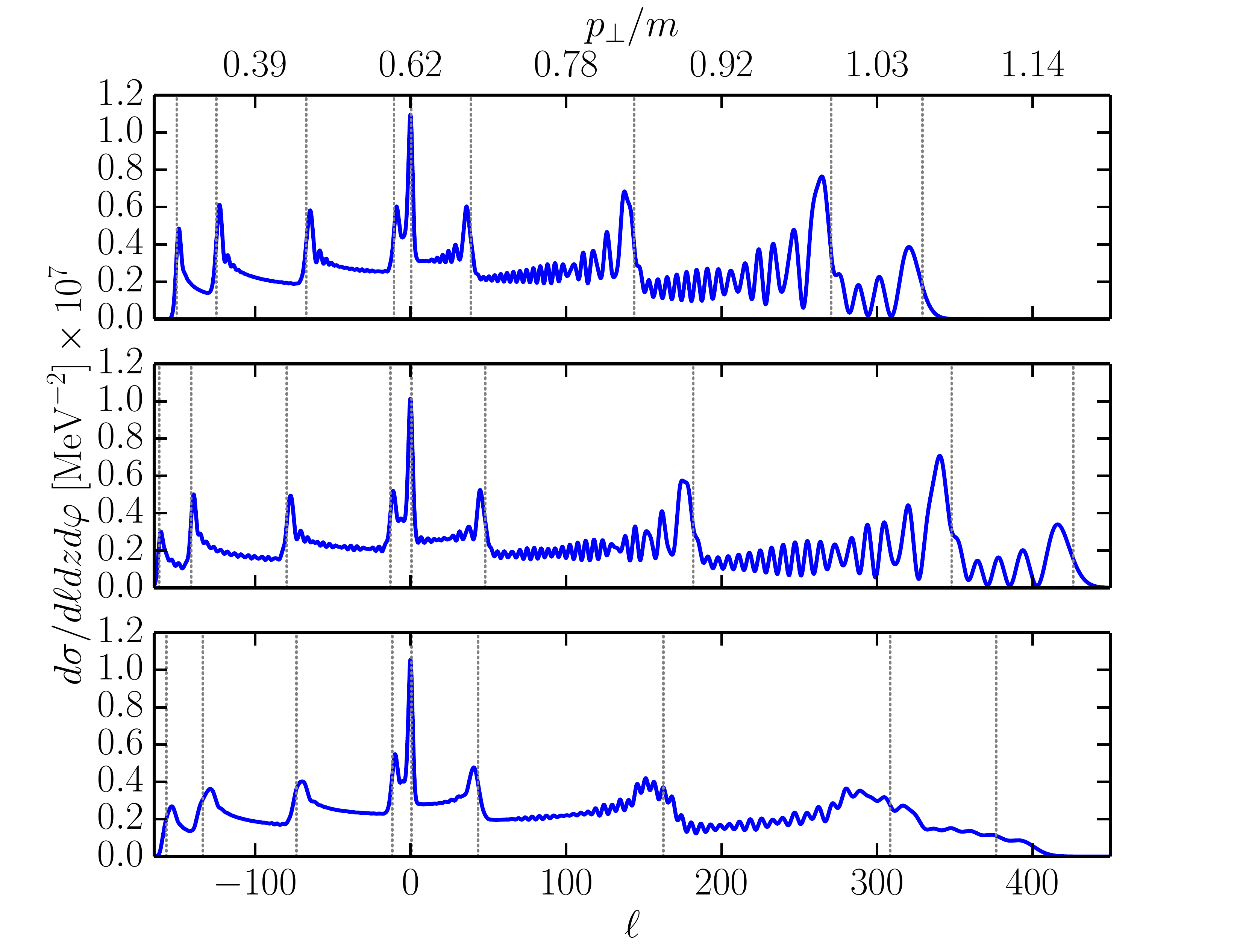}
\caption{As middle panel in Fig.~\ref{fig:1TN} but variation of $\gamma_L$
around $\gamma_L=2$.
Upper panel: $\gamma_L=2.22$, middle panel: $\gamma_L=1.82$, lower panel:
superposition of smoothed spectra for $\gamma_L=1.88\dots2.12$
corresponding to the laser intensity parameter $a_0=\gamma_L^{-1}=0.5\pm0.03$.}
\label{fig:2TN}
\end{figure}

The laser assisted, non-linear Breit-Wheeler process (cf.~\citep
{jansen_strongly_2013,jansen_strong-field_2015,wu_nonlinear_2014,
krajewska_breit-wheeler_2014,meuren_polarization-operator_2015}) is dealt with
within the strong-field QED (Furry picture) as reaction
$\gamma^\prime \to e^+_A + e^-_A$ where $e^\pm_A$ denote dressed
electron/positron states as Volkov solutions of the Dirac equation in a plane
wave model with vector potential of the common classical background field
\begin{equation}
A^\mu (\phi) = \gamma_X^{-1} f_X (\phi) \varepsilon^\mu_X \cos \phi
+ \gamma_L^{-1} f_L (\eta \phi) \varepsilon^\mu_L \cos \eta \phi,
\label{A_Tobias}
\end{equation}
where the polarization four-vectors are $\varepsilon^\mu_{X,L}$ and the
above defined Keldysh parameters $\gamma_{1,2}$ have been transposed to
$\gamma_{X,L}$; $\gamma'$ denotes the high-energy probe photon
traversing the field~\eqref{A_Tobias}.
The XFEL (frequency $\omega$) and laser (frequency $\eta\omega$, we
assume in the following $\eta\ll1$) beams are co-propagating and their linear polarizations are set perpendicular to each other to simplify the cumbersome numerical evaluation.
Both ones are pulsed as described, for the sake of computational
convenience, by the envelope functions
$f_X = \exp\{ - \phi^2 /(2 \tau_X^2)\}$ and $f_L = \cos^2 \left(\pi \phi /(2 \tau_L) \right)$
for $-\tau_L \le \phi \le + \tau_L$ and zero elsewhere for the latter pulse shape.
In contrast to~\eqref{A_AO} we treat here a somewhat more realistic
case with different pulse durations $\tau_X$ and $\tau_L$.
The invariant phase is $\phi = k \cdot x$
with the dot indicating the scalar product of the four-wave vector $k$ and the space-time
coordinate $x$.
It is convenient to parametrize the produced positron's
phase space by the following three variables: (i) the momentum
exchange parameter $\ell$, (ii) the azimuthal angle $\varphi$
with respect to the polarization direction of the assisting
laser field and (iii) the shifted rapidity
$z = \frac12 \log (p_+^+/p_+^-) + \frac 12 \log\left( (1 + \eta \ell) \omega_X / \omega_{X^\prime} \right)$.
The energy-momentum balance for laser assisted pair production can be put into
the form $k_{X^\prime}^\mu + k_X^\mu + \ell k_L^\mu = p_+^\mu + p_-^\mu$
($\mu$ is a Lorentz index, as above), where $\ell$ represents here an
hitherto unspecified momentum exchange between the assisting laser field $L$ and
the produced pair.
We define light-front coordinates, e.g. $x^\pm = x^0 \pm x^3$ and
$\vec x_\perp = (x_1, x_2)$
and analogously the light front
components of four-momenta of the probe photon $X^\prime$, the XFEL photon
$X$, the laser beam photons $L$ and positron (subscript $+$) and electron
(subscript $-$).
They become handy
because the laser four-momentum vectors only have one
non-vanishing light-front component $k_{X,L}^- = 2 \omega_{X,L}$.
In particular, the energy-momentum balance contains the three conservation equations
in light-front coordinates $k_{X^\prime}^+ = p_+^+ + p_-^+$ and
$\vec p_+^\perp = -\vec p_-^\perp$.
Moreover, the knowledge of all particle momenta
allows to calculate $\ell$ via the fourth equation
$\ell = \left( (p_+^- + p_-^-  - k_{X^\prime}^-) / k_X^- - 1) / \eta \right)$.
Treating $(\ell, z, \varphi)$ as independent variables the positron's four-momenta
are completely determined by the above energy-momentum balance equations,
see~\citep{nousch_spectral_2016} for details, in particular for
expressing the positron and electron momenta $p_\pm$ by $(\ell, z, \varphi)$.

The theoretical basis for formulating and evaluating the cross section
is outlined in~\citep{nousch_spectral_2016}. An example is displayed in the top panel of
Fig.~\ref{fig:1TN} for $\eta=1/600$, $\gamma_X=10^5$,
$\tau_X=7\tau/(4\pi\eta)$, $\gamma_L=2$, and $\tau_L=8\pi$ (examples for other parameters are
exhibited in~\citep{nousch_spectral_2016}) for kinematical conditions,
where the linear Breit-Wheeler effect for $X'+X$ is just above the threshold.
The involved spectral distribution (note that without the laser assistance only
the Breit-Wheeler peak centered at $\ell=0$ corresponding to $p_\perp=0.62\,m$
would appear
with a finite width as a consequence of the finite x ray pulse duration;
cf.~\citep{titov_enhanced_2012,titov_breit-wheeler_2013,nousch_pair_2012} for an
enhancement of pair production in short laser pulses).
The spectrum can be smoothed by a window function with a resolution scale of
$\delta=1.3$ (which is an ad hoc choice to better show the strength
distribution and which may be considered as a simple account for finite energy
resolution respective $p_\perp$ distribution) resulting in the red curve which
is exhibited separately in the middle panel.
In line with the interpretation in~\citep{nousch_spectral_2016,seipt_caustic_2015} the prominent peaks are caustics related to stationary phase points
determined by the turning points of the invariant phase $\phi$ as a function of
the variable $\ell$, see bottom panel.
This interpretation implies that the total cross section may be approximately
factorized into a plain Breit-Wheeler production part and a final-state
interaction part, where the latter one means the redistribution of the produced
particles by the impact of the laser field.
An analog interpretation of particle production in constant cross field
approximation in very strong fields have been put forward in~\citep
{meuren_semiclassical_2015}.
Figure~\ref{fig:3TN} demonstrates the strong impact of the laser field
intensity.
For smaller values of $\gamma_L$, the transverse momentum spectrum becomes more
stretched and its shape is changed.
This challenges the observability of the peaks related to caustics in multi-shot
experiments with fluctuating laser intensities.
In fact, for the unfavorable case of equally weighted deviations, a window of
less than $20\,\%$ is required to keep the peak structures, see
Fig.~\ref{fig:2TN}.
A truncated Gaussian distribution with $1\sigma$ width in the same interval is,
of course, much more favorable for keeping the peaks, in particular for larger
$p_\perp$.
We consider here only one particular case of the laser assisted,
linear Breit-Wheeler process which turns into the textbook Breit-Wheeler process
upon switching off the laser.
Non-linearities w.r.t.\ the XFEL beam, subthreshold (w.r.t.\ the $X'$ + XFEL
kinematics) effects combined with larger laser intensities, carrier envelope
phase effects, and a wider range of kinematical parameters (e.g.\ $\omega_L =
\mathcal O(\SI{1}{eV}$) need to be explored as well to arrive at a complete
picture.
Among the furthermore to be analyzed issues w.r.t.\ an experimental proposal are
non-monochromaticity and misalignment disturbances.

\section{Summary}
In summary we have supplied further important details of
(i) the amplification effect
of the assisted dynamical Schwinger effect and
(ii) the phase space redistribution in the laser assisted Breit-Wheeler process.
Both topics are motivated by the availability of x rays by XFELs
and upcoming ultra-high intensity laser beams. We consider the perspectives offered by the
combination of both beam types resulting in bi-frequent fields.
Concerning the Schwinger related investigations
we find that significant pair production by the dynamical assistance requires much
higher frequencies than such ones provided by XFEL beams in conjunction with
future ELI-IV field intensities.
The crucial challenge for the laser assisted Breit-Wheeler process and an access
to the predicted caustic structures is the high-energy probe photon beam in
combination with dedicated phase space selective detector set-ups.
The bi-frequent fields are dealt with as a classical background.
An avenue for further work is the proper account of quantum fluctuations and a
unifiying description of counter- and co-propagating fields.
\bigskip

\noindent\textbf{Acknowledgements}~~R. Sauerbrey, T. E. Cowan and H. Takabe are thanked for the collaboration within
the HIBEF project~\citep{hibef}.
D.B. and S.A.S. acknowledge support by NCN under grant number
UMO-2014/15/B/ST2/03752.\bigskip

\noindent Dedicated to the memory of Nikolay Borisovich Narozhny who
pioneered this field of research.

\bibliographystyle{jpp}
\bibliography{lit}

\begin{thebibliography}{50}
\expandafter\ifx\csname natexlab\endcsname\relax\def\natexlab#1{#1}\fi

\bibitem[Akal {\em et~al.\/}(2014)Akal, Villalba-Ch\'avez \&
  M{\"u}ller]{akal_electron-positron_2014}
{\sc Akal, I., Villalba-Ch\'avez, S. \& M{\"u}ller, C.} 2014 {Electron-positron
  pair production in a bifrequent oscillating electric field}. {\em Phys. Rev.
  D\/} {\bf 90}, 113004.

\bibitem[Augustin \& M{\"u}ller(2014)]{augustin_nonlinear_2014}
{\sc Augustin, S. \& M{\"u}ller, C.} 2014 {Nonlinear Bethe-Heitler Pair
  Creation in an Intense Two-Mode Laser Field}. {\em J. Phys.: Conf. Ser.\/}
  {\bf 497}~(1), 012020.

\bibitem[Bamber {\em et~al.\/}(1999)]{bamber_studies_1999}
{\sc Bamber, C. {\em et~al.\/}} 1999 {Studies of nonlinear QED in collisions of
  46.6-GeV electrons with intense laser pulses}. {\em Phys. Rev. D\/} {\bf 60},
  092004.

\bibitem[Bell \& Kirk(2008)]{bell_possibility_2008}
{\sc Bell, A. \& Kirk, J.} 2008 {Possibility of Prolific Pair Production with
  High-Power Lasers}. {\em Phys. Rev. Lett.\/} {\bf 101}~(20), 200403.

\bibitem[Bialynicki-Birula {\em et~al.\/}(1991)Bialynicki-Birula, Gornicki \&
  Rafelski]{bialynicki-birula_phase-space_1991}
{\sc Bialynicki-Birula, I., Gornicki, P. \& Rafelski, J.} 1991 {Phase space
  structure of the Dirac vacuum}. {\em Phys. Rev. D\/} {\bf 44}, 1825--1835.

\bibitem[Breit \& Wheeler(1934)]{breit_wheeler}
{\sc Breit, G. \& Wheeler, J.~A.} 1934 {Collision of Two Light Quanta}. {\em
  Phys. Rev.\/} {\bf 46}, 1087.

\bibitem[Brezin \& Itzykson(1970)]{brezin_pair_1970}
{\sc Brezin, E. \& Itzykson, C.} 1970 {Pair Production in Vacuum by an
  Alternating Field}. {\em Phys. Rev. D\/} {\bf 2}~(7), 1191--1199.

\bibitem[Burke {\em et~al.\/}(1997)]{burke_positron_1997}
{\sc Burke, D.~L. {\em et~al.\/}} 1997 {Positron Production in Multiphoton
  Light-by-Light Scattering}. {\em Phys. Rev. Lett.\/} {\bf 79}, 1626--1629.

\bibitem[Dabrowski \& Dunne(2014)]{dabrowski_super-adiabatic_2014}
{\sc Dabrowski, R. \& Dunne, G.~V.} 2014 {Superadiabatic particle number in
  Schwinger and de~Sitter particle production}. {\em Phys. Rev. D\/} {\bf
  90}~(2), 025021.

\bibitem[{Di Piazza} {\em et~al.\/}(2010){Di Piazza}, L{\"o}tstedt, Milstein \&
  Keitel]{di_piazza_effect_2010}
{\sc {Di Piazza}, A., L{\"o}tstedt, E., Milstein, A.~I. \& Keitel, C.~H.} 2010
  {Effect of a strong laser field on electron-positron photoproduction by
  relativistic nuclei}. {\em Phys. Rev. A\/} {\bf 81}~(6).

\bibitem[{Di Piazza} {\em et~al.\/}(2012){Di Piazza}, M{\"u}ller, Hatsagortsyan
  \& Keitel]{di_piazza_extremely_2012}
{\sc {Di Piazza}, A., M{\"u}ller, C., Hatsagortsyan, K.~Z. \& Keitel, C.~H.}
  2012 {Extremely high-intensity laser interactions with fundamental quantum
  systems}. {\em Rev.\ Mod.\ Phys.\/} {\bf 84}~(3), 1177--1228.

\bibitem[Dunne {\em et~al.\/}(2009)Dunne, Gies \&
  Sch{\"u}tzhold]{dunne_catalysis_2009}
{\sc Dunne, G.~V., Gies, H. \& Sch{\"u}tzhold, R.} 2009 {Catalysis of Schwinger
  vacuum pair production}. {\em Phys. Rev. D\/} {\bf 80}~(11), 111301.

\bibitem[{ELI}(2015)]{eli}
{\sc {ELI}} 2015 {European Extreme Light Infrastructure (ELI)}.
  \url{www.eli-laser.eu}.

\bibitem[Elkina {\em et~al.\/}(2011)Elkina, Fedotov, Kostyukov, Legkov,
  Narozhny, Nerush \& Ruhl]{elkina_qed_2011}
{\sc Elkina, N.~V., Fedotov, A.~M., Kostyukov, I.~Yu., Legkov, M.~V., Narozhny,
  N.~B., Nerush, E.~N. \& Ruhl, H.} 2011 {QED cascades induced by circularly
  polarized laser fields}. {\em Phys. Rev. {ST} Accel. Beams\/} {\bf 14}~(5),
  054401.

\bibitem[Gelis \& Tanji(2013)]{gelis_formulation_2013}
{\sc Gelis, F. \& Tanji, N.} 2013 {Formulation of the Schwinger mechanism in
  classical statistical field theory}. {\em Phys. Rev. D\/} {\bf 87}~(12),
  125035.

\bibitem[Gelis \& Tanji(2015)]{gelis_schwinger_2015}
{\sc Gelis, F. \& Tanji, N.} 2015 {Schwinger mechanism revisited}. {\em
  arXiv:1510.05451\/} .

\bibitem[H{\"a}hnel(2015)]{haehnel_bachelor_2015}
{\sc H{\"a}hnel, S.} 2015 {Paarerzeugung in elektrischen Feldern: Numerische
  Untersuchungen zum Schwinger-Effekt}. {Bachelor's thesis}, Technische
  Universit{\"a}t Dresden.

\bibitem[Hebenstreit(2011)]{hebenstreit_diss_2011}
{\sc Hebenstreit, F.} 2011 {Schwinger effect in inhomogeneous electric fields}.
  PhD thesis, Karl-Franzens-Universit{\"a}t Graz.

\bibitem[Hebenstreit \& Fillion-Gourdeau(2014)]{hebenstreit_optimization_2014}
{\sc Hebenstreit, F. \& Fillion-Gourdeau, F.} 2014 {Optimization of Schwinger
  pair production in colliding laser pulses}. {\em Phys. Lett. B\/} {\bf 739},
  189--195.

\bibitem[Heinz {\em et~al.\/}(2000)]{heinz_positron_2000}
{\sc Heinz, S. {\em et~al.\/}} 2000 {Positron spectra from internal pair
  conversion observed in U-238 + Ta-181 collisions}. {\em Eur. Phys. J. A\/}
  {\bf 9}, 55--61.

\bibitem[{HIBEF}(2015)]{hibef}
{\sc {HIBEF}} 2015 {The HIBEF project}. \url{www.hzdr.de/hgfbeamline}.

\bibitem[{HiPER}(2015)]{hiper}
{\sc {HiPER}} 2015 {High Power laser for Energy Research project (HiPER)}.
  \url{www.hiper-laser.org}.

\bibitem[Jansen \& M{\"u}ller(2013)]{jansen_strongly_2013}
{\sc Jansen, M. J.~A. \& M{\"u}ller, C.} 2013 {Strongly enhanced pair
  production in combined high- and low-frequency laser fields}. {\em Phys. Rev.
  A\/} {\bf 88}~(5), 052125.

\bibitem[Jansen \& M{\"u}ller(2015)]{jansen_strong-field_2015}
{\sc Jansen, M. J.~A. \& M{\"u}ller, C.} 2015 {Strong-Field Breit-Wheeler Pair
  Production in Short Laser Pulses: Identifying Multiphoton Interference and
  Carrier-Envelope Phase Effects}. {\em arXiv:1511.07660\/} .

\bibitem[King {\em et~al.\/}(2013)King, Elkina \& Ruhl]{king_photon_2013}
{\sc King, B., Elkina, N. \& Ruhl, H.} 2013 {Photon polarisation in
  electron-seeded pair-creation cascades}. {\em Phys. Rev. A\/} {\bf 87},
  042117.

\bibitem[Kohlf{\"u}rst {\em et~al.\/}(2014)Kohlf{\"u}rst, Gies \&
  Alkofer]{kohlfurst_effective_2014}
{\sc Kohlf{\"u}rst, C., Gies, H. \& Alkofer, R.} 2014 {Effective Mass
  Signatures in Multiphoton Pair Production}. {\em Phys. Rev. Lett.\/} {\bf
  112}~(5), 050402.

\bibitem[Kohlf{\"u}rst {\em et~al.\/}(2013)Kohlf{\"u}rst, Mitter, von Winckel,
  Hebenstreit \& Alkofer]{kohlfurst_optimizing_2013}
{\sc Kohlf{\"u}rst, C., Mitter, M., von Winckel, G., Hebenstreit, F. \&
  Alkofer, R.} 2013 {Optimizing the pulse shape for Schwinger pair production}.
  {\em Phys. Rev. D\/} {\bf 88}~(4), 045028.

\bibitem[Krajewska \& Kaminski(2014)]{krajewska_breit-wheeler_2014}
{\sc Krajewska, K. \& Kaminski, J.~Z.} 2014 {Breit-Wheeler pair creation by
  finite laser pulses}. {\em J. Phys. Conf. Ser.\/} {\bf 497}, 012016.

\bibitem[Meuren {\em et~al.\/}(2015{\natexlab{{\em a\/}}})Meuren,
  Hatsagortsyan, Keitel \& {Di Piazza}]{meuren_polarization-operator_2015}
{\sc Meuren, S., Hatsagortsyan, K.~Z., Keitel, C.~H. \& {Di Piazza}, A.}
  2015{\natexlab{{\em a\/}}} {Polarization-operator approach to pair creation
  in short laser pulses}. {\em Phys. Rev. D\/} {\bf 91}~(1), 013009.

\bibitem[Meuren {\em et~al.\/}(2015{\natexlab{{\em b\/}}})Meuren, Keitel \& {Di
  Piazza}]{meuren_semiclassical_2015}
{\sc Meuren, S., Keitel, C.~H. \& {Di Piazza}, A.} 2015{\natexlab{{\em b\/}}}
  {Semiclassical description of nonlinear electron-positron photoproduction in
  strong laser fields}. {\em arXiv:1503.03271\/} .

\bibitem[M{\"u}ller {\em et~al.\/}(1972)M{\"u}ller, Peitz, Rafelski \&
  Greiner]{muller_solution_1972}
{\sc M{\"u}ller, B., Peitz, H., Rafelski, J. \& Greiner, W.} 1972 {Solution of
  the Dirac equation for strong external fields}. {\em Phys. Rev. Lett.\/} {\bf
  28}, 1235.

\bibitem[M{\"u}ller {\em et~al.\/}(1973)M{\"u}ller, Rafelski \&
  Greiner]{muller_solution_1973}
{\sc M{\"u}ller, B., Rafelski, J. \& Greiner, W.} 1973 {Solution of the Dirac
  equation with two Coulomb centers}. {\em Phys. Lett. B\/} {\bf 47}, 5--7.

\bibitem[Narozhny {\em et~al.\/}(2004)Narozhny, Bulanov, Mur \&
  Popov]{narozhny_pair_2004}
{\sc Narozhny, N.~B., Bulanov, S.~S., Mur, V.~D. \& Popov, V.~S.} 2004
  {$e^+e^-$-pair production by a focused laser pulse in vacuum}. {\em Phys.
  Lett. A\/} {\bf 330}~(1-2), 1--6.

\bibitem[Narozhny \& Nikishov(1970)]{narozhny_the_1970}
{\sc Narozhny, N.~B. \& Nikishov, A.~I.} 1970 {The simplest processes in the
  pair creating electric field}. {\em Sov. J. Nucl. Phys.\/} {\bf 11}, 596.

\bibitem[Nousch {\em et~al.\/}(2012)Nousch, Seipt, K{\"a}mpfer \&
  Titov]{nousch_pair_2012}
{\sc Nousch, T., Seipt, D., K{\"a}mpfer, B. \& Titov, A.~I.} 2012 {Pair
  production in short laser pulses near threshold}. {\em Phys. Lett. B\/} {\bf
  715}~(1--3), 246--250.

\bibitem[Nousch {\em et~al.\/}(2016)Nousch, Seipt, K{\"a}mpfer \&
  Titov]{nousch_spectral_2016}
{\sc Nousch, T., Seipt, D., K{\"a}mpfer, B. \& Titov, A.~I.} 2016 {Spectral
  caustics in laser assisted Breit--Wheeler process}. {\em Phys. Lett. B\/} .

\bibitem[Otto {\em et~al.\/}(2015{\natexlab{{\em a\/}}})Otto, Seipt, Blaschke,
  K{\"a}mpfer \& Smolyansky]{otto_lifting_2015}
{\sc Otto, A., Seipt, D., Blaschke, D., K{\"a}mpfer, B. \& Smolyansky, S.~A.}
  2015{\natexlab{{\em a\/}}} {Lifting shell structures in the dynamically
  assisted Schwinger effect in periodic fields}. {\em Phys. Lett. B\/} {\bf
  740}, 335--340.

\bibitem[Otto {\em et~al.\/}(2015{\natexlab{{\em b\/}}})Otto, Seipt, Blaschke,
  Smolyansky \& K{\"a}mpfer]{otto_dynamical_2015}
{\sc Otto, A., Seipt, D., Blaschke, D.~B., Smolyansky, S.~A. \& K{\"a}mpfer,
  B.} 2015{\natexlab{{\em b\/}}} {Dynamical Schwinger process in a bifrequent
  electric field of finite duration: Survey on amplification}. {\em Phys.\
  Rev.\ D\/} {\bf 91}~(10), 105018.

\bibitem[Panferov {\em et~al.\/}(2015)Panferov, Smolyansky, Otto, K{\"a}mpfer,
  Blaschke \& Juchnowski]{panferov_assisted_2015}
{\sc Panferov, A.~D., Smolyansky, S.~A., Otto, A., K{\"a}mpfer, B., Blaschke,
  D.~B. \& Juchnowski, L.} 2015 {Assisted dynamical Schwinger effect: pair
  production in a pulsed bifrequent field}. {\em arXiv:1509.02901\/} .

\bibitem[Rafelski {\em et~al.\/}(1971)Rafelski, Fulcher \&
  Greiner]{rafelski_superheavy_1971}
{\sc Rafelski, J., Fulcher, L.~P. \& Greiner, W.} 1971 {Superheavy elements and
  an upper limit to the electric field strength}. {\em Phys. Rev. Lett.\/} {\bf
  27}, 958--961.

\bibitem[Rafelski {\em et~al.\/}(1978)Rafelski, M{\"u}ller \&
  Greiner]{greiner_3}
{\sc Rafelski, J., M{\"u}ller, B. \& Greiner, W.} 1978 {Spontaneous vacuum
  decay of supercritical nuclear composites}. {\em Zeitschrift f{\"u}r Physik A
  Atoms and Nuclei\/} {\bf 285}~(1), 49.

\bibitem[Sauter(1931)]{sauter}
{\sc Sauter, F.} 1931 {{\"U}ber} das {Verhalten} eines {Elektrons} im homogenen
  elektrischen {Feld} nach der relativistischen {Theorie} {Diracs}. {\em Z.
  Phys.\/} {\bf 69}~(11), 742.

\bibitem[Schmidt {\em et~al.\/}(1998)Schmidt, Blaschke, R{\"o}pke, Smolyansky,
  Prozorkevich \& Toneev]{schmidt_quantum_1998}
{\sc Schmidt, S.~M., Blaschke, D.~B., R{\"o}pke, G., Smolyansky, S.~A.,
  Prozorkevich, A.~V. \& Toneev, V.~D.} 1998 {A Quantum kinetic equation for
  particle production in the Schwinger mechanism}. {\em Int. J. Mod. Phys. E\/}
  {\bf 7}, 709.

\bibitem[Sch{\"u}tzhold {\em et~al.\/}(2008)Sch{\"u}tzhold, Gies \&
  Dunne]{schutzhold_dynamically_2008}
{\sc Sch{\"u}tzhold, R., Gies, H. \& Dunne, G.} 2008 {Dynamically Assisted
  Schwinger Mechanism}. {\em Phys. Rev. Lett.\/} {\bf 101}~(13), 130404.

\bibitem[Schwinger(1951)]{schwinger}
{\sc Schwinger, J.} 1951 {On Gauge Invariance and Vacuum Polarization}. {\em
  Phys. Rev.\/} {\bf 82}, 664.

\bibitem[Seipt {\em et~al.\/}(2015)Seipt, Surzhykov, Fritzsche \&
  K{\"a}mpfer]{seipt_caustic_2015}
{\sc Seipt, D., Surzhykov, A., Fritzsche, S. \& K{\"a}mpfer, B.} 2015 {Caustic
  structures in the spectrum of x-ray Compton scattering off electrons driven
  by a short intense laser pulse}. {\em arXiv:1507.08868\/} .

\bibitem[Titov {\em et~al.\/}(2013)Titov, K\"ampfer, Takabe \&
  Hosaka]{titov_breit-wheeler_2013}
{\sc Titov, A.~I., K\"ampfer, B., Takabe, H. \& Hosaka, A.} 2013 {Breit-Wheeler
  process in very short electromagnetic pulses}. {\em Phys. Rev. A\/} {\bf 87},
  042106.

\bibitem[Titov {\em et~al.\/}(2012)Titov, Takabe, K\"ampfer \&
  Hosaka]{titov_enhanced_2012}
{\sc Titov, A.~I., Takabe, H., K\"ampfer, B. \& Hosaka, A.} 2012 {Enhanced
  Subthreshold ${e}^{\mathbf{+}}{e}^{\mathbf{-}}$ Production in Short Laser
  Pulses}. {\em Phys. Rev. Lett.\/} {\bf 108}, 240406.

\bibitem[Wu \& Xue(2014)]{wu_nonlinear_2014}
{\sc Wu, Y.~B. \& Xue, S.~S.} 2014 {Nonlinear Breit-Wheeler process in the
  collision of a photon with two plane waves}. {\em Phys. Rev. D\/} {\bf
  90}~(1), 013009.

\bibitem[Zou {\em et~al.\/}(2015)]{zou_design_2015}
{\sc Zou, J.~P. {\em et~al.\/}} 2015 {Design and current progress of the
  Apollon 10 PW project}. {\em HPLaser\/} {\bf 3}.

\end{thebibliography}
\end{document}